\documentstyle[sprocl,epsfig]{article}
\begin{document}
\title{BCP3: Summary of Theory}
\author{R. D. Peccei}
\address{Department of Physics and Astronomy, University of
California\\
Los Angeles, California 90095}

\maketitle

\begin{abstract}

\noindent I discuss a number of the highlights in theory presented at
the BCP3 conference.  These included new, and more stringent,
CKM fits; a critical overview of heavy hadron lifetimes; 
progress in computing rates and CP-asymmetries in charmless
B-decays; a thorough discussion of the implications of the
new results on $\epsilon^{\prime}/\epsilon$; and,
finally, a peek at the future, trying to estimate how well
one is going to be able to measure the unitarity triangle angles.
\end{abstract}

\noindent It is, as usual, very difficult to summarize the myriad of
talks in a conference dedicated to a forefront topic in
particle physics.  BCP3 was no exception.  Thus here, asking the
indulgence of the speakers whose work I will not mention, I will
only touch on a few of the highlights in theory which
I thought were particularly interesting.

\section{CKM Fits:  Where Do We Stand?}

During BCP3 there was much discussion of how well we can
determine the parameters in the CKM matrix.  In practice,
this means finding the allowed region, at a certain
confidence level, in the $\rho-\eta$ plane - with $\rho$
and $\eta$ being the usual Wolfenstein \cite{Wolf}
parameters.  The ingredients of the global fits discussed
here particularly by Stocchi \cite{Stocchi}, as well as by
Kim \cite{Kim} and Eigen \cite{Eigen}, are crucially
dependent on the errors one assigns to the values one infers experimentally
for $\vert V_{ub}\vert$ and $\vert V_{cb}\vert$.
~\cite{Thaler}~\cite{Jin}  Although the data on $\sin 2\beta$
coming from the Tevatron \cite{Trischuk}, and now also
from ALEPH,~ \cite{Wu}~\cite{Forty} is not accurate enough to
have much impact on the allowed region in the $\rho - \eta$
plane, the sharpened LEP bound on $\Delta m_s$, presented
here by Willocq,~ \cite{Willocq} provides a significant
restriction.

When considering the result of the CKM fits, it is
important to distinguish three different ingredients that
enter into these fits

\begin{description}

\item{i)} There are a number of experimental inputs which
have quite {\it negligible experimental errors}.  These
include: the CP-violating parameter $\vert \epsilon\vert$; the $B_d - \bar{B}_d$ mass
difference,
$\Delta m_d$; the
$\Delta m_s$ bound mentioned above;~ \cite{Willocq} and the value of the sine
of the Cabibbo angle, $\sin\theta_C\equiv \lambda$.

\item{ii)} There are a number of parameters which have, what one
may call {\it controllable
errors}.  These include  $\vert V_{cb}\vert$ ; the
ratio $\vert V_{ub}\vert / \vert V_{cb} \vert$ and the
$SU(3)$-breaking parameter
$\xi=f_{B_{s}} {\sqrt{B_{B_{s}}}} /f_{B_{d}} {\sqrt{B_{B_{d}}}}$.

\item{iii)} There
are also
associated theoretical parameters, whose errors are due to {\it theoretical
uncertaintes} in determining appropriate hadronic matrix elements.  These
include  $B_K$, the parameter which measures how different the matrix
element $< K\vert ({\bar{d}}s)^2_{V-A}\vert {\bar{K}}>$ is from the vacuum
saturation result $(B_K=1)$; and the analogous parameter for the
$B_d$ system, $B_d f_{B_{d}}^2$, which is connected to the
matrix element $<B_d \vert ({\bar{d}} b)^2_{V-A} \vert {\bar{B}}_d>$.

\end{description}

The parameters with {\it{controllable errors}}, in principle, are the ones which
can be improved using more experimental information.~\footnote{I will discuss
explicitly below  how one can effect this
error reduction in the case of $\vert V_{cb}\vert$.  Similar ideas can be brought to bear on
$\vert V_{ub}\vert / \vert V_{cb}\vert$ and also on $\xi$.}  In
contrast, it is difficult to reduce, or even correctly estimate, the
theoretical errors associated with parameters like $B_K$, because they
are connected with the way one attempts to calculate the hadronic matrix
elements.  Although Soni \cite{Soni} in this meeting has given 15\% errors
on $B_K$ and $f_{B_{d}} {\sqrt{B_{B_d}}}$~$ [B_K =
0.85\pm 0.13$;~$ f_{B_{d}}{\sqrt{B_{B_{d}}}}= (230 \pm
30)MeV]$,
the systematic error on each of these quantities is much harder to pin
down, since it depends on the
uncertainties associated with going from a quenched to a fully unquenched
calculation of the relevant
matrix elements.

The fit of Stocchi shown in Fig. 1 uses the parameters detailed in Table 1.
Although I find this set
of parameters (and the fit!) perfectly reasonable, I believe that one
should not take the 95\%
confidence limit region shown in Fig. 1 strictly as such.  There is enough
"theoretical error"
in this whole business that the most prudent approach is to transform a
"formal" 95\%
confidence limit region into an "effective" 68\% limit region.  Even after doing
such an unorthodox
thing, one cannot but be impressed by how well the CKM model fits the data!

\begin{figure}[t]
\center
\epsfig{file=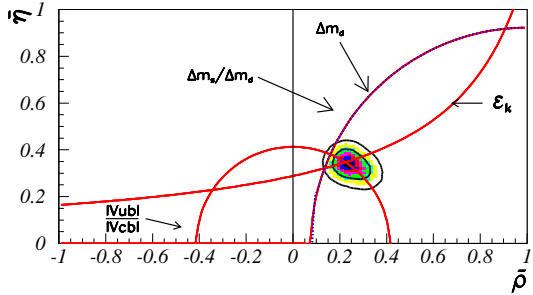,width=3.5in}
\caption{CKM fit presented in Ref [2].}
\end{figure}

\begin{table}
\caption{ Input parameters for the fit in Fig. 1, from reference [2]}
\begin{center}
\begin{tabular}{|c|c|} \hline
Parameter & Value\\ \hline
$\vert \epsilon \vert$ & $(2.280 \pm 0.019)\times 10^{-3}$ \\ \hline
$\vert V_{ub}\vert / \vert V_{cb}\vert$ & $0.080 \pm 0.006 \pm 0.016
$~~Cornell \\ \hline
$\vert V_{ub}\vert / \vert V_{cb}\vert$ & $0.104 \pm 0.012 \pm 0.015$ ~~LEP \\ \hline
$\Delta m_d$ & $(0.472 \pm 0.016)~{\rm ps}^{-1}$  \\ \hline
$\Delta m_s$ & $>14.3 ~{\rm ps}^{-1}$\\ \hline
$\vert V_{cb}\vert$ & $(40.4 \pm 1.9)\times 10^{-3}$ \\ \hline
$B_K$ & $0.86 \pm 0.06 \pm 0.08$ \\ \hline
$f_{B_d}{\sqrt{B_{B_{d}}}}$& $(210 \pm 29 \pm 12)
$MeV \\ \hline
$\xi$ & $1.11 \pm 0.02 \pm 0.05$\\ \hline 
\end{tabular}
\end{center}
\end{table}

As I mentioned earlier, the new result from LEP reported here by
Willocq~\cite{Willocq} on the $B_s - \bar{B}_s $ mass
difference is of considerable importance for the final
$\rho-\eta$ plane fit.  Last year this parameter was bounded a
bit more weakly
$(\Delta m_s >12.4~ {\rm ps^{-1}})$ so that the final $\rho - \eta$ plane
fit permitted values of $\rho < 0$.~ \cite{RDPSA}  With the
present bound, however, one cannot contemplate any longer negative
values for $\rho$--- a point emphasized by Deshpande \cite{Desh} at
this meeting.  If $\rho >0$, it follows that the CKM CP-violating
phase   $\gamma$ cannot be as large as $90^{0}$.
In turn, this implies
that one cannot any longer contemplate the superweak solution for the
unitarity triangle angles
$\alpha$ and $\beta$, in which $\sin 2 \beta = \sin 2 \alpha$.  Indeed, the fit
of the CKM parameters
presented by Stocchi here leads to a best value for $\sin 2 \beta$ and $\sin 2
\alpha$ which are quite
different from each other:~ \cite{Stocchi}
\begin{equation}
\sin 2\beta = 0.75 \pm 0.06:~~ \sin 2 \alpha = -0.22 \pm 0.24~.
\end{equation}

I remark, that the 95\% confidence limit on $\sin2 \beta$ which one infers from
the CKM analysis,
\begin{equation}
0.62 \leq \sin 2 \beta \leq 0.88,
\end{equation}
is precisely in the range determined by present experiments.  The value
reported here for $\sin 2
\beta$, obtained by ALEPH ~\cite{Wu} ~\cite {Forty}
 \begin{equation}
\sin 2 \beta = 0.93^{+0.64 ~+ 0.36}_{-0.88 ~ -0.24}~,
\end{equation}
when combined with the CDF result \cite{Trischuk} [$\sin 2 \beta = 0.79^{+
0.41}_{-0.44}$], leads to an average value
\begin{equation}
< \sin 2 \beta > = 0.91 \pm 0.35~,
\end{equation}
which is perfectly compatible with the range for $\sin 2 \beta$ obtained indirectly from the CKM analysis.
Furthermore, this experimental result, by itself, is already almost
significant statistically.

It is possible to imagine improving considerably the allowed value for $\sin
2 \beta$  and
other unitarity triangle parameters by improving the {\it controllable
errors} on $\vert
V_{ub}\vert / \vert V_{cb}\vert$ and on the other parameters which enter in the CKM
analysis.
Let me illustrate how these improvements can come about by focusing
specifically on the case of
$\vert V_{cb}\vert$.  If one could trust the parton model, and
one knew the
mass of the b-quark precisely, one could directly extract $\vert
V_{cb}\vert$ from the semileptonic
width for b-quarks to decay into final states containing c-quarks, given by the standard formula:

\begin{equation}
\Gamma (b \to c \ell {\bar{\nu}}_e) = \vert V_{cb} \vert^2
\frac{G^2_F}{192 \pi^3} m^5_b~. 
\end{equation}
This formula, however, is of no use since $m_b$ itself is not
known precisly enough and, furthermore, there are corrections to
the parton model! 

What one does, in practice, is to use the
Heavy Quark Effective Theory (HQET) to replace $m_b$ by the
mass of the B-meson.  In so doing the rate (5) above is
corrected by the addition of terms involving the matrix elements
of the operators $0_i$ which enter in the operator produced
expansion underlying the HQET.~ \cite{HQET}  To $0(1/m_b^2)$
one has two new operators contributing to the total width, which are
characterized by two parameters $\lambda_1$, and $\lambda_2$
given by
\begin{eqnarray}
\lambda_1&=& \frac{1}{2M_B}<B\vert{\bar{b}}(iD)^2
b \vert B >\\ \lambda_2 &=& \frac{g_3}{2M_B} <
B \vert {\bar{b}} \sigma_{\mu \nu} G^{\mu \nu} b \vert B>~,\nonumber
\end{eqnarray}
where $G^{\mu\nu }$ is the gluon field strength tensor.  In addition, to this
order in the
heavy quark expansion,~ \cite{HQET} the Fermi momentum ${\bar{\Lambda}}$ enters
in the
formulas relating the mass of the b-quark to that of the $B$ and
$B^*$
mesons:

\begin{eqnarray}
M_{B} &=&m_b  + {\bar{\Lambda}}-\frac{ (\lambda_1 + 3 \lambda_2)}{m_b} \\
M_{B}^{*} &=&m_b + {\bar{\Lambda}} - \frac{(\lambda_1 -
\lambda_2)}{m_b}~.\nonumber
\end{eqnarray}
Using the physical masses for $M_B$ and $M_{B}^*$ yields directly a value for
$\lambda_2 ~[ \lambda_2 \simeq 0.12~ {\rm GeV}^2]$.

Using the heavy quark expansion one can obtain a formula for the
semileptonic rate
$\Gamma(B \to X_c \ell \bar{\nu}_\ell)$ which involves $M_{B}$ rather than $m_b$,
at the price of
some corrections involving $\lambda_1, \lambda_2$, and ${\bar{\Lambda}}$. One
finds, including
leading order QCD effects, \cite{Falk} the formula:

\begin{eqnarray}
\Gamma (B \to X_c \ell {\bar{\nu}}_\ell) &=& \vert V_{cb}\vert^2
\frac{G^2_F}{192 \pi^3}
M_{B}^5 \Phi (\frac{M_D}{M_B}) [1 - 1.54\frac{\alpha_s}{\pi}+ ...]\\
&\cdot&  \{1 - 1.65 \frac{{\bar{\Lambda}}}{M_{B}}\ - 3.18
\frac{\lambda_1}{M_B^2} + 0.02 \frac{\lambda_2}{M_B^2} +
...\}~.\nonumber
\end{eqnarray}
Here $\Phi (M_D/M_{B})$ is a phase space factor,
while the first square bracket contains the $0
(\alpha_s)$ perturbative QCD corrections.  The terms in the
curly brackets are the dominant corrections coming from the
heavy quark expansion.  If, besides $\lambda_2$, one knew $\lambda_1$ and
${\bar{\Lambda}}$  with good accuracy,
one could determine $\vert V_{cb} \vert$ from the
semileptonic width with a small $0 ( 1/m_b^3)$ theory
error.  The parameters $\lambda_1, \lambda_2$, and ${\bar{\Lambda}}$ are, in principle, {\it
controllable} since they can be inferred from
properties of the lepton spectrum and of the hadronic mass spectrum in non-leptonic  $B$-decays
.~\cite{Falk} Specifically, moments of the lepton spectrum and/or the
hadronic mass spectrum can be used to determine
$\lambda_1$ and ${\bar{\Lambda}}$.

\begin{figure}[t]
\center
\epsfig{file=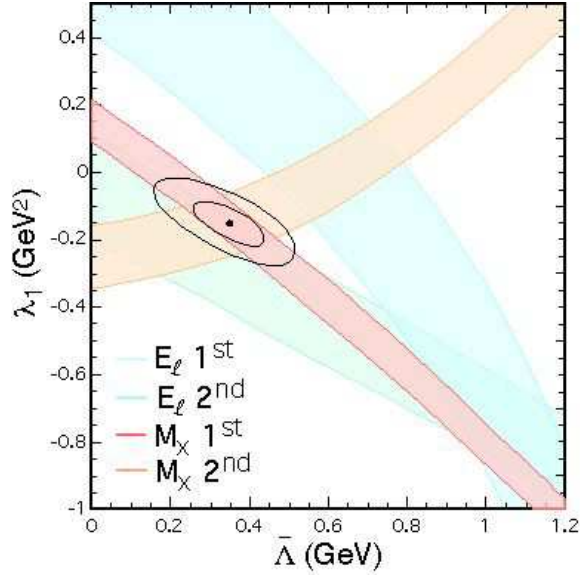,height=3in}
\caption{Preliminary analysis by the CLEO Collaboration of the moments of the hadronic mass spectrum (dark shaded) and moments of the leptonic energy spectrum (light shaded) to determine $\lambda_1$ and $\bar{\Lambda}$.}
\end{figure}

Unfortunately, as shown in Fig. 2, a preliminary analysis
by the CLEO collaboration --- presented here by Thaler
\cite {Thaler} --- give inconsistent results. The values for 
$\lambda_1$ and $ {\bar{\Lambda}}$ determined from the leptonic
spectrum do not seem to agree with those computed from an
analysis of the moments of the hadronic mass spectrum.  In my
view, the latter values $[\lambda_1 = (-0.13 \pm 0.01 \pm 0.06)~
{\rm GeV^2}$;~ ${\bar{\Lambda}} = (0.33 \pm 0.02 \pm 0.08)$~ GeV] are
probably more trustworthy, since they are less sensitive to some
of the necessary experimental cuts.  Indeed, these values are
quite close to the values reported by W. M. Zhang \cite{Zhang} at
this meeting, coming from a B-bound state analysis using an
explicit phenomenological wavefunction for the B-meson.  At
any rate, using these techniques, it is likely that one will
be able to eventually reduce the theoretical error on $\vert
V_{cb}\vert$ to about $1 \times 10^{-3}~ (2.5\%)$ and on $\vert
V_{ub}\vert$ to about $ 4\times 10^{-4}~ (10\%)$, with a
corresponding reduction of the allowed region in the $\rho-\eta$
plane.  As Misiak \cite{Misiak} discussed in BCP3, this same
approach can be used in other contexts --- in particular, to better
constrain the dependence on the photon energy of the branching
ratio for $B \to X_s \gamma$ above some minimum photon energy cut.

\section{Lifetimes of Heavy Hadrons}

There is reasonable theoretical understanding of the specific
pattern of lifetimes of heavy hadrons containing $c$ and $b$
quarks.  This topic was reviewed here by Bigi ~\cite{Bigi} and
specific aspects were considered by Melic ~\cite{Melic} and by Yang
~\cite{Yang}.  However, as the data on these lifetimes is now
rather precise,~ \cite{Ukagawa} a puzzle has emerged.  I want
to discuss this briefly here.

One can understand the total decay width -- and hence the lifetime
-- of a heavy hadron theoretically via the HQET.  The total width
is given by the discontinuity of the correlator of the weak
Hamiltonian responsible for the decay with itself.  This
discontinuity is given by a sum of matrix elements of
operators of increasing dimensonality, arising in an operator
product expansion of the correlator.  Schematically, one finds
in this way the formula~\cite{Bigi}

\begin{eqnarray}
\Gamma_{\rm tot} \sim G_F^2 m_Q^5 &\left[ <H _Q|\bar{Q} Q
| H_Q> +
{\frac{C_g}{m_Q^2}}<H_Q
\vert \bar{Q} \sigma_{\mu \nu} G^{\mu \nu} Q \vert H_Q > \right. \\
 &\left. +{\frac{C_{4f}}{m_Q^3} }< H_Q \vert \bar{Q} \Gamma q
\bar{q} \Gamma Q \vert H_Q > + ... \right] ~, \nonumber
\end{eqnarray}
where $m_Q$ is the mass of the heavy quark.  As
can be gleaned from the above, the expansion parameter
for the total width is of order
$\Lambda_{QCD}/m_Q$.  Hence one expects that the effects of the
non-leading terms in Eq. (9) should be bigger in charm decays than
in bottom decays.  Indeed, experimentally $\tau(D^+)/\tau(D^o)\sim
2.5$ while the $B^+$ and $B^{o}_d$ lifetimes are nearly equal
$[\tau (B^+)/\tau (B^o_d) = 1.066 \pm 0.024$ ~\cite{Ukagawa}]

More specifically, the situation regarding the decays of charmed
hadrons seems rather satisfactory.~ \cite{Bigi}  The dominant
effect that causes the large difference between the lifetimes of
the $D^+$ and $D^0$ mesons can be traced to the Pauli interference
~\cite{Guberina} originating from the 4-fermion term in Eq. (9).
These 4-fermion terms are also crucial to explain the pattern of
charmed baryon lifetimes, as detailed by Melic ~\cite{Melic} at
this meeting.  The fact that $\tau (D_s)/\tau(D^o)$ is near to
1.2 ~$[\tau (D^o) = 408 \pm 4.1~ {\rm fs} ;~ \tau (D_s) = 486 \pm 15
~{\rm fs}$~\cite{Ukagawa}] is probably evidence for some weak
annihilation contribution $(c{\bar{s}} \to  W^+ )$  in the
$D_s$ width.~ \cite{Bigi}  If so, perhaps one should be able to
see some evidence for glueball or $\eta^{\prime}$ decays of the
$D_s$, processes like $D_s \to $ glueball $\ell {\bar{\nu}}_\ell$ ; $D_s \to
\eta^{\prime} \ell {\bar{\nu}}_\ell$.

Although the prediction that the lifetime differences in the
$B$-sector should be small is well borne out by the data,
there is a nagging problem with the $\Lambda_b$ lifetime.~
\cite{Bigi}  The Pauli interference effects suppress the
lifetime of the  $\Lambda_b$ relative to that of the
$B_d$-meson and one can write
\begin{equation}
\frac{\tau (\Lambda_b)}{\tau(B_d)} = 1 - \Delta
\end{equation}
Bigi~\cite{Bigi} estimates $\Delta \simeq 0.05$, while Melic
~\cite{Melic} can push $\Delta$ to perhaps $\Delta \simeq
0.1$ by taking a very light $b$-quark mass.  However, experimentally
this ratio is closer to 0.8, with the latest results giving
$\tau (\Lambda_b)/\tau(B_d) = 0.79 \pm 0.05$.~
\cite{Wu}  It is possible that this signals a real discrepancy between
theory and experiment that needs explaining.  However, in my view,
given the fine agreement of all the other heavy hadron lifetimes with
the predictions coming from HQET, perhaps one should not worry overmuch at this stage  about a (2-3) $\sigma$ discrepancy.  Hopefully, future
data from the Tevatron should help resolve this issue.

\section{Charmless $B$-decays:  Rates and CP-Asymmetries}

At BCP3 there was lots of discussion of 2-body charmless
$B$-decay.  Data on these decays was presented by J. Smith
~\cite{Smith} and J. Alexander ~\cite{Alexander}, while various
theoretical aspects were touched upon by H-Y Cheng ~\cite{Cheng}; C. D.
Lu~\cite{Lu}, H.-n. Li~\cite{Li}, M. Suzuki ~\cite{Suzuki}and C.
Bhattacharya. ~\cite{Bhattacharya}

H.- Y. Cheng~\cite{Cheng} in particular, summarized the recent
considerable progress achieved in calculating within QCD the relative rates for the decays of $B$-meson into 2 pseudoscalar states $[B \to PP]$,
2 vector states $[B \to VV]$ and a vector and a pseudoscalar
state $[B\to VP]$ .  The present treatment formalizes
(and, in a sense, justifies and explains) the old diagrammatic
approach to this problem of Ling Lie Chau.~ \cite{Chau}  The basic idea
of these calculations is well known.  One starts with an effective
weak Hamiltonian given by a sum of operators

\begin{equation}
 H_{\rm eff} =\sum_{i} C_i (\mu) 0_i(\mu)~,
\end{equation}
 and one tries to estimate the matrix elements of
these operators by using factorization. However, then one correct this
procedure by including (some) non-factorizable pieces.

Technically, when one uses factorization,  in the process of
splitting up the operators as a product of currents
the scale dependence of the
matrix elements of $0_i (\mu)$ is not preserved:
\begin{equation}
<M_1 M_2 \vert 0_i (\mu) \vert B> = <M_1 \vert  J_i\vert B>
<M_2 \vert J_i \vert B >~.
\end{equation}
However, one can reorganize the Wilson coefficient expansion
~\cite{CLY} to effectively recover $\mu$-independent coefficients
$C^{\rm eff}_i$ - at least to $0(\alpha_s)$.  For Penguin operators
these effects are rather large, amounting to about a  $ 50\%$ increase for
$C^{\rm eff}_6$.~\cite{Cheng}  Non-factorizable contributions are
incorporated as $1/N_c$ corrections.  As Cheng~\cite{Cheng} discussed, the recent advance comes from
understanding that, depending on which operator $0_i$ one is
considering, the correction factors have different $N_c$-factors associated with them .  This has been justified in
the heavy quark limit recently by Beneke, Buchalla, Neubert and
Sachrajda. \cite{BBNS}  In
particular, for $(V-A)(V-A)$ operators $N^{\rm eff}_c \simeq 2$, while
for $(V+A)(V-A)$ operators $N_c^{\rm eff}\simeq 6$.

H.- Y. Cheng~\cite{Cheng} compared the theoretical predictions resulting
from the above considerations with experimental results.  The
overall comparison is quite good, but there are  two extant
problems.  These are:
\begin{description}
\item{i)} It is difficult to get the branching ratio $BR(B \to KK)
\simeq 0.25 BR (B \to K\pi)$ as is observed experimentally.
Typically, one finds much larger $B\to \pi \pi$ branching fractions
than are observed.

\item{ii)} It is very hard to push the $BR(B\to \eta^{\prime} K)$ to
the very large level observed $(BR \sim 80 \times 10^{-6})$  Typically, theoretically 
one can reach at most $BR(B\to \eta^{\prime}K)\sim 50 \times
10^{-6}$.
In fact, to get to values as large as this, it is necessary that $N^{\rm eff}_c =2$ for $(V-A)(V-A)$ operators, as is suggested theoretically.
\end{description}

J. Smith~\cite{Smith} discussed a, theoretically less sophisticated,
fit of the measured charmless $B$-decays. In this fit, which was carried out by  Hou, Smith
and W$\ddot{\rm u}$rtheim, ~\cite{HSW} one simply considers
the amplitude for the B-decays in question to be given by a Tree plus a Penguin
contribution, neglecting throughout any strong rescattering phases.
Further, these authors  use (naive) factorization to calculate the relevant
matrix elements.  Since the Tree amplitudes depended on the CKM phase
$\gamma$~$ [ T = \vert T \vert e^{-i\gamma}]$, this fit determines this
phase.  Remarkably the fit discussed by Smith~\cite{Smith} determines
$\gamma$ quite accurately:  
\begin{equation}
\gamma = (114^{+25}_{-21})^o~.
\end{equation}
 However, it is difficult to judge the
reliability of the approach.  Furthermore, also in this case the
branching fraction for the decay $B\to \eta^{\prime}K$ is too low
~$[BR(B\to \eta^{\prime}K)\sim 30 \times 10^{-6}]$.

The situation regarding {\it direct CP-asymmetries} is much less clear,
both experimentally and theoretically,  Experimentally, as Alexander
~\cite{Alexander} indicated, one is statistics limited so that $\delta
A_{\rm CP viol.}$ is typically of $0(0.2)$.  For example, one has
~\cite{Alexander}
\begin{eqnarray}
A_{\rm CP viol.}(K^\pm \pi^\mp) &=& - 0.04 \pm 0.16 \\
A_{\rm CP viol.}(K^0 \pi^\pm) &=& +0.18 \pm 0.24 \nonumber
\end{eqnarray}
Improvements will scale as $1/\sqrt{N}$ and one will need an integrated
luminosity of $0(100~ {\rm fb}^{-1})$ to
get to $\delta A_{\rm{CP viol.}}\simeq 0.04!$

Theoretically to be able to predict these direct CP-asymmetries one
needs a good estimate of the strong rescattering phases.  If one
writes for the amplitudes of two charge -conjugate processes
\begin{equation}
T = T_1 + T_2 e^{i\delta_w} e^{i\delta_s} ;~~
{\bar{T}} = T_1 + T_2 e^{-i \delta_w} e^{i\delta_s}~,
\end{equation}
one sees that $A_{{\rm CP viol.}}$ vanishes if there is no strong rescattering
phase $\delta_s$ between the two amplitudes:

\begin{equation}
 A_{{\rm CP viol.}} = \frac{2r \sin \delta_w \sin \delta_s}{1 + r^2  + 2r \cos
\delta_w \cos \delta_s}~,
\end{equation}
where $r = T_1/T_2$.  For sizeable effects, in addition to having a non-negligible rescattering phase, one
needs a relatively large weak CP-violating phase $\delta_w$ (which is probably OK in the
CKM model) and a ratio $r \sim 1$.  Since, typically, the
two amplitudes involved are Penguin and Tree amplitudes,
one needs these amplitudes to be comparable in size and to have a
large rescattering phase between them ---  not too likely a
possibility!  Numerically, for example, if $r = 0.25, \delta_w\equiv\gamma
= 60^o$ and $\delta_s = 30^o$, one obtains $A_{{\rm CP viol.}} \simeq
0.1$.

At BCP3 three different approaches were discussed to try to
estimate the rescattering phases $\delta_s$ which one might
expect.  All three approaches have some inherent
difficulties, demonstrating how challenging really it is to
have a reliable estimate for $\delta_s$.  C. D. Lu~\cite{Lu}
used the old idea of Bander, Silverman and Soni~\cite{BSS} to
extract a rescattering phase simply from the phase associated
with the discontinuity of Penguin graphs.  This phase depends
on the momentum transfer carried by the gluon, $\delta_s
(k^2)$.  However, since the relevant $k^2$ are rather low, it
is natural to question whether one can trust the
discontinuity calculated in a pure quark picture to such
values of $k^2$.  In contrast, H.-n. Li~\cite{Li} in his talk
at BCP3 estimated $\delta_s$ as the rescattering phase which
emerges in the Brodsky-Lepage bound state formalism
~\cite{BLP} by means of factorization.  In this case, the question is
whether one can really use these techniques given the large
energy release in the decay process $B\to M_1 M_2$.

Mahiko Suzuki~\cite{Suzuki} presented a more general
discussion of the problem based on hadronic methods.  He
argued, I believe correctly, that it is really
difficult to apply perturbative QCD ideas --- even if one
includes some resummation --- for the process at hand, since
the effective scale is only of order $k_{\rm eff} \sim
\sqrt{\Lambda_{QCD}m_b} \sim 1.5$ GeV.  However, Suzuki
also pointed out that the rescattering phase $\delta_s$ is also
difficult to estimate by hadronic methods.  This is because
the phase associated with, say, the $\vert K \pi>$ final
state is not simply the phase associated with elastic
$K\pi$ scattering. For the energies in question
\begin{equation}
\vert K\pi> = \sum_f e^{i\delta_f} \vert \alpha_f>~,
 \end{equation}
and most of the states that contribute in the above sum are inelastic
(roughtly 80\%). Given this fact, Suzuki tried to estimate the effective
phase $\delta_s$ that emerges by using a random phase
approximation.  Using this approximation, he was able  to correlate the
resulting phase $\delta_s$ with the elasticity of the process, with the result depending on
how favored or disfavored is the factorization of the
amplitude.  When factorization is not favoreda $\delta_s$ is
bigger.  However, if one really has a large rescattering phase
$\delta_s$ there are also sizable distorsions of the
amplitude --- a result which Suzuki~\cite{Suzuki} points out was
first understood by Fermi! Basically, not only is there a
change in the imaginary part associated with the amplitude,
but also the real part is affected:
\begin{equation}
T \to T {\rm exp}\left[\frac {P}{\pi}\int \frac{ds^{\prime}
\delta_s(s^{\prime})}{s^{\prime} - s}\right]~.
\end{equation}
These considerations make it unlikely that one will be able to really ever get
a reliable theoretical prediction for a direct CP-asymmetry.  However, this fact
should not discourage experimentalists from looking for such CP-asymmetries.

 \section{Imputs from the Strange Quark Sector}

In the last year, the new results on $\epsilon^{\prime}/\epsilon$ announced by
KTeV~\cite{KTeV} and by the NA48 Collaboration~\cite{NA48} have
generated an enormous amount of interest.  Not surprisingly, this
subject was also a topic of considerable discussion at BCP3.

\subsection {\protect $\epsilon^{\prime}/\epsilon$ Results and their
Interpretation}

In BCP3 the experimental situation regarding $\epsilon^{\prime}/\epsilon$
was reviewed by A. Roodman~\cite{Roodman} and by A. Nappi~\cite{Nappi}, while
the theoretical aspets of these recent results were discussed by
Bertolini,~\cite{Bertolini} Soni,~\cite{Soni} Masiero,~\cite{Masiero} and
Soldan.~\cite{Soldan} In my view, the nicest aspect of the new KTEV/NA48
results is that they resolve the discrepancy that existed betwen the old
Fermilab result, coming from the E731 experiment,~\cite{E731} and the results of the old CERN experiment NA31.~ \cite{NA31}  The
present results for $\epsilon^{\prime}/\epsilon$:
\begin{equation}
{\rm Re}~  \epsilon^{\prime}/\epsilon =\left\{\begin{array}{c}
(28.0 \pm 4.1) \times 10^{-4}~~{\rm Ref~ [35]}\\
(18.5 \pm  7.3) \times 10^{-4}~~
{\rm Ref~ [36]}\end{array}
\right.
\end{equation}
when combined with the E731 and NA31 results lead to a world average value
for this quantity:
\begin{equation}
<{\rm Re} ~\epsilon^{\prime}/\epsilon>=(21.2 \pm 4.6) \times 10^{-4}~,
\end{equation}
which clearly establishes the existence of $\Delta S = 1$ CP-violation.
This result provides the first {\em direct} confirmation that $\sin \gamma
\not=0 $ or, equivalently, that $\eta \not=0$--- something which could only be inferred from the CKM
analysis we discussed earlier.

This said, however, it is difficult to extract from the
present measurement of $\epsilon^{\prime}/\epsilon$ a
precise value for $\eta$. This is because the value
of\ $\epsilon^{\prime}/\epsilon$,  although proportional to
$\eta$, is rather uncertain due to uncertainties in the
calculation of the relevant hadronic matrix element.  This
point was emphasized by Bertolini~\cite{Bertolini} at this
meeting. His argument can be illustrated by making use of an
approximate formula for $\epsilon^{\prime}/\epsilon$ due to Buras and his collaborators.~\cite{Buras} As is well known, both matrix elements of gluonic Penguin operators
and of electroweak Penguin operators contribute to the
process
$s\to d {\bar{q}} q$.  Although the gluonic Penguin
contributions are of $0(\alpha_s)$, and hence enhanced with respect to the electroweak Penguin contributions which are of $O(\alpha)$, they are
suppressed by the $\Delta I = 1/2$ rule
since they are proportional to $A_2/A_0$.~ \cite{BB}  The
electroweak Penguin contributions, on the other hand, although naturally
small do not suffer from the $\Delta
I = 1/2$ suppresion and, furthermore, are
enhanced by a factor of
$m_{t}^2/M_W^2$.~\cite{FR} As a result, for
$\epsilon^{\prime}/\epsilon$ both Penguin operator
contributions are of a similar magnitude and,
because they enter with opposite sign, they render
the theoretical value for this  quantity rather
uncertain.

This can be appreciated quite nicely from the
approximate formula for
$\epsilon^{\prime}/\epsilon$ derived by
Buras and his collaborators.~\cite{Buras} One has
\begin{equation}
\epsilon^{\prime}/\epsilon \simeq 34 \eta \left[B_6^{1/2}
- 0.53 B_{8}^{1/2} \right] \left(\frac{110 MeV}{m_s(2GeV)}\right)^2
\times 10^{-4}~.
\end{equation}
Here $B_6$ and $B_{8}$ are the matrix elements of
the gluonic Penguin operator and of the electroweak
Penguin operator, respectively --- normalized so that
in the vacuum insertion approximation $B_6 =
B_{8} = 1$.  Because, as we have seen, $\eta
\simeq 0.3$, the above formula in this
approximation gives $\epsilon^{\prime}/\epsilon
\sim 5 \times 10^{-4}$, which is much below the value 
measured by KTeV and NA48.  Indeed, to get agreement with
the present world average [cf Eq. (20)], one needs to
stretch all the parameters in Buras's formula.  Namely,
one needs to maximize $\eta$  [ $\eta \simeq 0.4]$; decrease
the value one assumes for $m_s (2GeV)$ [perhaps to $m_s$
as low as $m_s (2GeV) \simeq 90$ MeV]; increase
$B_6^{1/2}$ from unity --- something which is not clear  one can obtain in
lattice calculations,~ \cite{Martinelli} but which appears to be true
in the chiral quark model~\cite{Bertolini}; and decrease
$B_{8 }^{1/2}$ --- something which emerges naturally both in lattice
and $1/N$ calculations.~\cite{Paschos}

In this meeting, Bertolini~\cite{Bertolini} emphasized that
what improves the {\em post-dictions} of
$\epsilon^{\prime}/\epsilon$ is to incorporate in the
calculation the trend observed in the chiral quark model and
in the $1/N$ approximation that $B_6^{1/2}/B_8^{1/2} \simeq
2$, and not unity as expected in the vacuum insertion
approximation.  However, this is just a phenomenological
observation and one will not really trust a theoretical
result for  $\epsilon^{\prime}/\epsilon $ until the
lattice results quantitatively  arrive at a value
for $\epsilon^{\prime}/\epsilon$. Unfortunately, at the moment, there is considerable controversy on what this value might be. This was clear from Soni's talk ~\cite{Soni} at BCP3. 

Soni ~\cite{Soni} argued forcefully  that the extant lattice calculations for
$\epsilon^{\prime}/\epsilon$ which use staggered fermions
are not to be trusted, since they have trouble
correctly incorporating the needed counterterms and
cannot account well for chiral mixing.  However, when one
uses domain wall fermions, which have the correct chiral
behaviour, one arrives at a result for
$\epsilon^{\prime}/\epsilon$  which is
difficult to believe:~\cite{Soni}
\begin{equation}
[\epsilon^{\prime}/\epsilon]_{DWF} = (-120 \pm 60) \times
10^{-4}~.
\end{equation}
Not only is the sign of $\epsilon^{\prime}/\epsilon$
reversed from what one measures experimentally --- due to a
large negative contribution from the, so called,
"eye diagrams"~ \cite{Soni} --- but the overall
magnitude is also rather large.

Given the theoretical disarray concerning
$\epsilon^{\prime}/\epsilon$, it appears to me premature
to try to invoke the presence of some new physics to
$\lq\lq$explain" the experimental value of
$\epsilon^{\prime}/\epsilon$. However, considering new physics
with {\em correlated} predictions is interesting.  For example,
as discussed by Masiero ~\cite{Masiero} here, in supersymmetric
models with a large $\delta_{RL}$ phase, not only does one get
$\epsilon^{\prime}/\epsilon$ to be large, but one also predicts
largish values for the electron dipole moment and the lepton
flavor violating process $\mu \to e \gamma$.  As Pakvasa
~\cite{Pakvasa} observed, such models also tend to give a rather
large CP-violating  contribution in $\Lambda$-decays.

\subsection{Rare K-decays}

In contrast to $\epsilon^{\prime}/\epsilon$, as D'Ambrosio
~\cite{DA} discussed in BCP3, there are rare K-decays where the
theoretical prediction are on much firmer ground.  In
particular, the "golden  modes" $K_L\to \pi^0 \nu
{\bar{\nu }}$ (whose branching ratio is proportional to
$\eta^2$) and $K^{+}\to \pi^+ \nu {\bar{\nu}}$ (whose
branching ratio is proportional to $\vert V_{td}\vert^2$) have
theoretical errors of order 1\% and 5\%, respectively.  The
Brookhaven experiment BNL E787 has observed one event
corresponding to this latter process and one infers a branching
ratio~\cite{Kettel}
 \begin{equation}
BR (K^+ \to\pi^+ \nu {\bar{\nu}}) =(1.5^{+3.5}_{-1.3})\times 10^{-10}~,  
\end{equation}
which is totally consistent with the expectations of the standard model
$[(BR)_{SM} = (0.82 \pm 0.32)\times 10^{-10}]$.  There are strong hopes
~\cite{Kettel} that the proposed BNL experiment E949 will be able to
actually pin down a value for $\vert V_{td}\vert$, since they expect of
order 10 events at the SM level of sensitivity.

Clearly, since the branching ratio for $K_L \to \pi^0 \nu {\bar{\nu}}$
is directly proportional to $\eta^2$ and it has a negligible theoretical
error, a measurement of this process would have a significant impact on
our knowledge of the CKM matrix.  Unfortunately, the decay $K_L \to
\pi^0 \nu {\bar{\nu}}$ is extremely challenging experimentally, since
one expects a very small branching ratio in the  SM $[(BR)_{SM} \simeq 3
\times 10^{-11}]$  and, further, one has an all neutral final state.
Nevertheless, as discussed here by Hsiung ~\cite{Hsiung} there are a
number of proposed experiments in a planning stage, both at Brookhaven
[KOPIO], Fermilab [KAMI] and KEK [E391].  However, to determine $\eta$ at
a significant level (with error of order $\delta \eta/ \eta \sim 0.1$)
one needs to measure the $K_L \to \pi^0 \nu {\bar{\nu}}$ branching
ratio to an accuracy of order 20\%.  This is very hard indeed!

As Pakvasa \cite{Pakvasa} emphasized in his talk at BCP3, the theoretical
predictions for CP-violation in Hyperon decays are likely to have less
uncertainty than those for $\epsilon^{\prime}/\epsilon$. However, in Hyperon decays, at least in the standard model,
the expectations for CP-violating phenomena are very small and appear to
be well below the present experimental reach.~\cite{KBL}  Typically, one
expects theoretically for $\Lambda$-decays  a CP-violating contribution of order  $A_{\rm CP viol.}(\Lambda)\sim
10^{-5}$, while experimentally one can reach only a level of order
$A_{\rm CP viol.}(\Lambda)\vert_{exp} \sim 10^{-3}-10^{-4}$.

\section{Looking at the Future:  Determing the Phases in the Unitarity
Triangle}

The most interesting challenge of the B-factories now entering into
operation, and of future collider B experiments, is to try to pin down
the angles in the unitarity triangle.  As Trischuk ~\cite{Trischuk}
discussed at this meeting for the most favorable mode, involving the
decay $B_d \to \psi K_s$, one can expect to reach the level $\delta \sin 2\beta
\simeq 0.1$ at the B-factories, with an integrated luminosity of $50 ~{\rm fb}^{-1}$. To reach this same level of accuracy at the Tevatron,
one probably will need about $2~{\rm fb}^{-1}$ of integrated luminosity.  These luminosity numbers are likely to be achieved in the
next five
years.  Although the "uncertainty" $\delta \sin 2 \beta \simeq 0.1$ is
precisely that which obtains from present day CKM fits,~ \cite{Stocchi} the
results
on $\sin 2 \beta$ both at the B-factories and at the Tevatron will involve
an {\em
actual} measurement of a CP-violating asymmetry.  Thus they are extremely
important--- even if they probably will not significantly improve the knowledge of $\sin 2 \beta$ obtained indirectly through CKM fits.

In the same vein, it is also important to check that the value of $\sin 2 \beta$
measured in a variety of different physical processes is in fact the same.
Indeed, as remarked by a number of people at this meeting
,~\cite{Okada}~\cite{He} this may well be the best way to look for
physics beyond the standard model.  For instance, if there were to be an
extra $b
\to s$ Penguin phase $\Phi_P$, the process $B_d \to \psi K_s$ would still approximately measure $\sin 2
\beta$, while the process $B_d \to \Phi K_s$ (which is Penguin dominated) would
measure $\sin (2 \beta + \Phi_P)$.

Extracting the two other angles in the unitarity triangle, $\alpha$ and $\gamma$,
from experiment is likely to be much more challenging.  This will require
measuring many processes, as Sheldon Stone ~\cite{Stone} emphasized in his nice
overview at BCP3. Furthermore, one will need to use, at the same time, theory input rather
judiciously.~ \cite{Desh}~\cite{Gronau} That this is so can be appreciated
in a number of ways.  For example, as is well known,~ \cite{GL} the process $B_d
\to \pi^+\pi^-$ does not measure purely $\sin 2 \alpha$ since there is likely
significant Penguin pollution.  In principle, one could imagine estimating
these effects by studying $B_d \to \pi^o \pi^o$.~ \cite{GL}  However, this decay
is estimated to have a $BR \sim 10^{-7}$, which is too small to make this an
effective means to control Penguin pollution.  Deshpande ~\cite{Desh}
at this meeting suggested what may be a useful alternative. Namely, using a theoretical calculation
to extract $\alpha$ from the value $\alpha_{\rm meaus}$ gotten experimentally from the
$B_d \to \pi^+\pi^- $ asymmetry.  The model calculations he presented appeared
rather encouraging.

Gronau ~\cite{Gronau} discussed, analogously, how theoretical input --- in this case
the $SU(3)$ transformation properties of the weak Hamiltonian --- can be used to
constrain $\gamma$ for the $\Delta S = 1$  $B \to K \pi$ process.
The piece of the decay amplitude which transforms as the ${\overline{15}}$
dimensional representation in the Tree amplitude essentially fixes the
electroweak-Penguin amplitude
\begin{equation}
T_{ EWP} ({\overline{15}})= \delta_{EW} T_{\rm
tree}({\overline{15}})~,
\end{equation}
with $\delta_{EW} = 0.65 \pm 0.15$ a calculable number in the
standard
model.  Then using that $\vert K^o \pi^+>+ \sqrt{2}\vert K^+ \pi^o>$
 transforms as the 27 representation one can obtain useful interelations among $B \to K \pi$ process, from
which one
can extract $\gamma$.  These techniques can be used also in other contexts and
Gronau ~\cite{Gronau} estimated that they may lead to a determination of $\gamma$
with an
error $\delta\gamma = \pm 20^o$.

Obviously, with new data coming in from the B-factories, and soon also from the
Tevatron, we should expect interesting news for BCP4!

\section*{Acknowledgements}

I would like to thank both George Hou and Hai-Yang Cheng for their splendid
hospitality in Taipei.  This work was supported in part by the Department of
Energy under Contract DE-FE03-91ER40662, Task C.

\end{document}